\documentclass[letter]{aa}
\usepackage{graphicx}
\usepackage{txfonts}
\usepackage{natbib}
\newcommand\bb[1] {   \mbox{\boldmath{$#1$}}  }

\begin{document}

\title{MHD simulations of the magnetorotational instability in a
  shearing box with zero net flux: the case $Pm=4$}
\author{S\'ebastien Fromang\inst{1,2}}

\offprints{S.Fromang}

\institute{CEA, Irfu, SAp, Centre de Saclay, F-91191 Gif-sur-Yvette,
  France \and UMR AIM, CEA-CNRS-Univ. Paris VII, Centre de Saclay,
  F-91191 Gif-sur-Yvette, France. \\ \email{sebastien.fromang@cea.fr}}

\date{Accepted; Received; in original form;}

\label{firstpage}

\abstract
{}
{This letter investigates the transport properties of MHD turbulence 
induced by the magnetorotational instability at large Reynolds numbers
$Re$ when the magnetic Prandtl number $Pm$ is larger than unity.}
{Three MHD simulations of the magnetorotational instability (MRI)
in the unstratified shearing box with zero net flux are
presented. These simulations are performed with the code Zeus and
consider the evolution of the rate of angular momentum transport
as $Re$ is gradually increased from $3125$ to $12500$ while
simultaneously keeping 
$Pm=4$. To ensure that the small scale features of the flow 
are well resolved, the resolution varies from $128$ cells per disk 
scaleheight to $512$ cells per scaleheight. The latter constitutes the
highest resolution of an MRI turbulence simulation to date.}
{The rate of angular momentum transport, measured using the $\alpha$
parameter, depends only very weakly on the Reynolds
number: $\alpha$ is found to be about $7 \times 10^{-3}$ with
variations around this mean value bounded by $15\%$ in all
simulations. There is no systematic evolution with
$Re$. For the best resolved model, the kinetic   
energy power spectrum tentatively displays a power-law range with
an exponent $-3/2$, while the magnetic energy is found to shift to
smaller and smaller scales as the magnetic Reynolds number 
increases. A couple of different diagnostics both suggest a
well--defined injection length of a fraction of a scaleheight.}
{The results presented in this letter are consistent with the MRI being
able to transport angular momentum efficiently at large Reynolds
numbers when $Pm=4$ in unstratified zero net flux shearing boxes.}

\authorrunning{S.Fromang}
\titlerunning{High resolution simulations of the MRI}
\maketitle

\section{Introduction} 
\label{intro}

Angular momentum transport in accretion disks has been an outstanding
issue in theoretical astrophysics for decades. To date the most
likely mechanism appears to be MHD turbulence driven by the
magnetorotational instability
\citep[MRI,][]{balbus&hawley91,balbus&hawley98}. Several numerical 
simulations have been performed to study its 
properties. The most popular approach is to work in the local approximation,
using the shearing box model, as pioneered by 
\citet{hawleyetal95}, \citet{hawleyetal96} or
\citet{brandenburgetal95}. These early simulations have shown
that MRI--powered MHD turbulence is a robust mechanism that transports
angular momentum outward. The rate of transport, measured by the
famous $\alpha$ parameter \citep{shakura&sunyaev73} depends on the
field geometry but is always positive, indicating outward flux of
angular momentum. The results obtained in the 1990's however obviously suffered
from the limited computational resources available at that time. With
no mean vertical magnetic field threading the shearing 
box (a field geometry referred to
as {\it the zero net flux} case), \citet{fromang&pap07} recently 
demonstrated with the code Zeus \citep{hawley&stone95} that it is
indeed a problem: $\alpha$ decreases by a factor of two each time the
resolution is doubled. 
This behavior has since been shown to be very
  robust as it has been confirmed by simulations performed with codes
using different algorithms \citep{simonetal09,guanetal09}.
This
result, although it raised the concern that MRI--induced transport could
vanish at infinite resolution, was interpreted as an indication that
the small scale 
behavior of the flow is an important ingredient to determine
the rate of MRI--induced angular momentum transport: small scale
explicit dissipation coefficients, namely viscosity and resistivity,
need to be included in the simulations. With such
calculations \citet{lesur&longaretti07} showed that, for a nonzero
vertical mean magnetic field, $\alpha$ rises with the 
magnetic Prandtl number $Pm$, the ratio of viscosity over
resistivity. This result is actually very general: it is
independent of the field geometry and was also found for a mean
toroidal magnetic field \citep{simon&hawley09} and 
in the zero net flux case of interest here
\citep{fromangetal07}. Recently \citet{simonetal09} measured the {\it 
  numerical} dissipation properties of the code Athena
\citep{gardiner&stone08,stoneetal08}. They found that an increase in
resolution amounts to an increase of the {\it numerical} Reynolds
numbers, while keeping the {\it effective} 
magnetic Prandtl number (i.e. the ratio between the numerical
viscosity and the numerical resistivity) roughly constant 
and equal to about two. In light of these results a possible
interpretation of the findings of \citet{fromang&pap07} is that
$\alpha$ is decreasing when the {\it physical} Reynolds number
increases at fixed $Pm$. If 
unchecked, this decreasing $\alpha$ would mean that MRI--induced MHD
turbulence is ineffective at transporting angular momentum 
without a mean flux, even in systems that have $Pm$ values
higher than unity. 
Here, high resolution numerical simulations in which $Re$ and $Rm$ are
simultaneously increased while keeping their ratio $Pm$ constant are
used to examine if this is indeed the case.

%The plan of the letter is as follows. In section~\ref{basic_prop}, the
%numerical setup of the simulations is
%described. In section~\ref{flow_prop_sec}, the properties of the
%turbulence are analyzed. Three statistical diagnostics are
%considered: the rate of angular momentum transport, the kinetic and
%magnetic energy power spectrum and the two points correlation
%function. Finally, the conclusions and consequences of the
%results are discussed in section~\ref{conclusion_section}. 

\section{Numerical setup}
\label{basic_prop}

%\begin{table*}[t]
%\begin{center}\begin{tabular}{@{}ccccccc}\hline\hline
%Model & Run duration & Resolution & Reynolds number & $\alpha_{Rey}$ & $\alpha_{Max}$ & $\alpha$ \\
%\hline\hline
%$Re3125$ & 150 orbits & $(128,192,128)$ & 3125 & $1.4 \times 10^{-3} \pm 4.2 \times 10^{-4}$ & $6.6 \times 10^{-3} \pm 1.5 \times 10^{-3}$  & $7.9 \times 10^{-3} \pm 1.8 \times 10^{-3}$ \\
%$Re6250$ & 90 orbits & $(256,384,256)$ & 6250 & $1.0 \times 10^{-3} \pm 2.2 \times 10^{-4}$ & $4.8 \times 10^{-3} \pm 5.7 \times 10^{-4}$ & $5.9 \times 10^{-3} \pm 6.7 \times 10^{-4}$ \\
%$Re12500$ & 45 orbits & $(512,768,512)$ & 12500 & $1.4 \times 10^{-3} \pm 3.4 \times 10^{-4}$ & $7.0 \times 10^{-3} \pm 1.2 \times 10^{-3}$ & $8.4 \times 10^{-3} \pm 1.5 \times 10^{-3}$ \\
%\hline\hline
%\end{tabular}
%\caption{Properties of the simulations and time averaged values of the
%transport coefficients. All runs share the same magnetic Prandtl
%number $Pm=4$.}
%\label{model_prop}
%\end{center}
%\end{table*}

\begin{table}[t]
\begin{center}\begin{tabular}{@{}cccc}\hline\hline
Model & Resolution & $Re$ & $\alpha$ \\
\hline\hline
$Re3125$ & $(128,192,128)$ & 3125 & $7.9 \times 10^{-3} \pm 4.5 \times 10^{-4}$ \\
$Re6250$ & $(256,384,256)$ & 6250 & $5.9 \times 10^{-3} \pm 1.8 \times 10^{-4}$ \\
$Re12500$ & $(512,768,512)$ & 12500 & $8.4 \times 10^{-3} \pm 3.8 \times 10^{-4}$ \\
\hline\hline
\end{tabular}
\caption{Properties of the simulations and time averaged value of
  $\alpha$. The errors on $\alpha$ are computed
following \citet{longaretti&lesur10}: the time history of $\alpha$ is divided in 
N bins of size $\tau$. $\tau$ is varied between $0.1$ and $8$
orbits. For each N, the standard deviation $\sigma_N$ is computed
according to $\sigma_N=[\Sigma (\alpha_i-\alpha)/N]^{1/2}$, where
$\alpha_i$ is the mean value in bin i. For large N, $\sigma_N$ scales
like $N^{-1/2}$. The errors reported on $\alpha$ use that scaling to
estimate $\sigma_N$ when $\tau=40$ orbits, the time duration over
which the mean values of $\alpha$ are calculated.
%The errors on $\alpha$ are computed as described by
%\citet{longaretti&lesur10}. The time history of $\alpha$ is divided in
%N bins of size $\tau$. $\tau$ is varied between $0.1$ and $8$
%orbits. For each N, the standard deviation $\sigma_N$ is computed
%according to $\sigma_N=[\Sigma (\alpha_i-\alpha)/N]^{1/2}$, where
%$\alpha_i$ is the mean value in bin i. The error reported in column
%three is the extrapolation of $\sigma_N$ to $\tau \sim 40$ orbits.
}
\label{model_prop}
\end{center}
\end{table}

%\begin{figure*}
%\begin{center}
%\includegraphics[scale=0.25]{figure/by_128_pm4.ps}
%\includegraphics[scale=0.25]{figure/by_256_pm4.ps}
%\includegraphics[scale=0.25]{figure/by_512_pm4.ps}
%%\includegraphics[scale=0.25]{figure/vz_128_pm4.ps}
%%\includegraphics[scale=0.25]{figure/vz_256_pm4.ps}
%%\includegraphics[scale=0.25]{figure/vz_512_pm4.ps}
%\caption{Typical snapshots of the azimuthal component of the magnetic
%  field in the $(x,z)$ plane for models $Re3125$ ({\it left panel}),
%  $Re6250$ ({\it middle panel}) and $Re12500$ ({\it right panel}).}
%\label{pm4_snapshots}
%\end{center}
%\end{figure*}

%128 run:
% save_30 -> time=94.5 orbits
% save_41 -> time=127.8 orbits
% save_43 -> time=132.6 orbits
% save_47 -> time=144.0 orbits
% save_49 -> time=149.4 orbits
%256 run:
% save_20 -> time=86.8 orbits
% save_39 -> time=133.4 orbits
%512 run:
% save_10 -> time=105.1 orbits
% save_30 -> time=128.0 orbits

In the simulations described below, the non--ideal
MHD equations (i.e. including viscosity $\nu$ and resistivity $\eta$)
are solved in the unstratified shearing box 
\citep{goldreich&lyndenbell65} by the
code Zeus \citep{hawley&stone95}. The setup is identical to that used
by \citet{fromangetal07}: the shearing box rotates around the central
point mass with angular velocity $\Omega$ (thus defining the orbital
time $T_{orb}=2\pi / \Omega$), the equation of state is
isothermal with the sound speed $c_0$, and the size of the box is fixed to
$(L_x,L_y,L_z)=(H,\pi H,H)$, where $H=c_0/\Omega$ is the disk
scaleheight. As mentioned in the introduction, the magnetic flux
threading the disk vanishes in all directions. 
%In addition, the magnetic flux threading the disk
%vanishes in either directions: this is imposed by superposing the
%initial form for the magnetic field $\bb{B}$ to the background
%equilibrium at $t=0$ as
%\begin{equation}
%\bb{B}=B_0 \sin (2\pi x/H) \, ,
%\end{equation}
%where $B_0$ is chosen such that the volume averaged value of $\beta$,
%the ratio of the thermal pressure $P$ to the magnetic pressure $P_{mag}=B^2/2$
%equals $400$. The MRI is trigerred by adding small random velocity
%fluctuations to that state.
%For a complete determination of the flow properties, $\nu$ and $\eta$
%need to be specified. In this letter, the structure of the flow is
%studied when both coefficients are simulatenously decreased while
%keeping their ratio fixed to $Pm=\nu/\eta=4$. 
Three simulations are presented here. They share the same
value for the magnetic Prandtl number $Pm=\nu/\eta=4$. The Reynolds
number $Re=c_0H/\nu$ is gradually increased from $Re=3125$ (hereafter
labeled model 
$Re3125$) to $Re=6250$ (model $Re6250$) and finally $Re=12500$ (model
$Re12500$). The resolution is increased at the same time as the
Reynolds number to ensure that the smallest scale features of the flow
are always resolved. Model $Re3125$ is identical to model
$128Re3125Pm4$ of \citet{fromangetal07}, for which different
diagnostics have shown that $128$ cells per scaleheight are sufficient
when using Zeus. Thus the resolutions $(N_x,N_y,N_z)=(128,192,128)$,
$(256,384,256)$ and  $(512,768,512)$ are adopted respectively for
model $Re3125$, $Re6250$ and 
$Re12500$\footnote{Model $Re12500$, with $512$ cells per scaleheight,
constitutes the highest resolution published so far of MRI induced
turbulence. With about $2 \times 10^8$ cells, the simulation required over
$1.4$ million timesteps to be completed and a total of about $350000$
CPU hours on the CEA supercomputer BULL Novascale 3045 hosted in
France by CCRT.
% on a cluster composed of Intanium2 1.6 GHz available at the
%CCRT supercomputing center
}. 
For model $Re12500$, it was found that early transients associated with the
linear instability kept affecting the flow for long times,
resulting in prohibitively long simulations. For the computational
cost of that simulation to remain acceptable, the
following procedure was used: model $Re3125$ was run from $t=0$ to
$t=150$ orbits, starting from the initial state described above and
identical to that used by \citet{fromangetal07}. At $t=60$, the flow
was interpolated on a grid twice finer. The dissipation coefficients
were reduced by a factor of two and the model was restarted between
$t=60$ and $t=150$ orbits. This constitutes model $Re6250$. This
procedure was repeated at time $t=90$ orbits to produce model
$Re12500$. The latter was run between $t=90$ and $t=135$ orbits. 
The properties of the three models are summarized in
Table~\ref{model_prop}: the first column gives the label of the model,
the second column reports its resolution $(N_x,N_y,N_z)$ and the third
the Reynolds number $Re$ for that run. All models share the
same value $Pm=4$. Finally, the last column in Table~\ref{model_prop}
gives time--averaged values of $\alpha$ that are discussed in the
subsections below.
%The properties of the three models are summarized in
%table~\ref{model_prop}: the first column gives the label of the model,
%the second column report the duration of the simulation, the third
%column its resolution $(N_x,N_y,N_z)$ and the fourth column the
%Reynolds number $Re$ for that run. All models share the same value
%$Pm=\nu/\eta=4$ for the magnetic Prandtl number. The following three
%columns in table~\ref{model_prop} gives time averaged values of the
%transport coefficients discussed in the following
%subsections. 

\section{Flow properties}
\label{flow_prop_sec}

%In order to illustrate the wealth of information contained in the
%highest resolution simulation, snapshots of model $Re12500$ are
%presented in figure~\ref{pm4_snapshots}. From left to right, the
%different panels represent the density, the vertical velocity and the
%toroidal component of the magnetic field in the meridional plane of
%the disk. 
In the three models flow features typical of
unstratified shearing boxes simulations are recovered: weakly 
non--axisymmetric density waves propagate radially in the box 
\citep{heinemann&papaloizou09a,heinemann&papaloizou09b} 
on top of smaller scales velocity and magnetic field turbulent
fluctuations, the latter exhibiting a tangled structure typical of
$Pm$ values higher than unity \citep{schekochihinetal04}. Below we
concentrate on the transport properties of the 
turbulence, the shape of the kinetic and magnetic energy power spectra
and the the two points correlation function.

\subsection{Angular momentum transport}

\begin{figure}
\begin{center}
\includegraphics[scale=0.38]{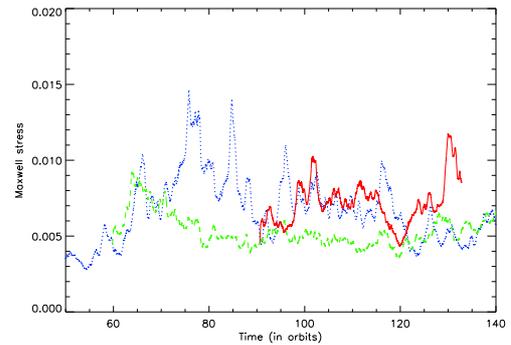}
\caption{Time history of the Maxwell stress tensor for model $Re3125$
  ({\it blue dotted line}), $Re6250$ ({\it green dashed line}) and $Re12500$ ({\it
    red solid line}). The three curves are consistent with the same time
  averaged value for $\alpha_{Max}$, independently of the Reynolds number.}
\label{maxwell_pm4}
\end{center}
\end{figure}

The angular momentum transport properties of the turbulence in the
three models are assessed by calculating the $\alpha$ parameter,
the sum of the Reynolds stress tensor $\alpha_{Rey}$ and the Maxwell
stress tensor $\alpha_{Max}$. All three coefficients are calculated as
in \citet{fromang&pap07}. The time history of $\alpha_{Max}$ is shown
in Fig.~\ref{maxwell_pm4} for models $Re3125$, $Re6250$ and
$Re12500$ respectively, using a dotted, a dashed and a solid line. The
result is dramatically different from the results of
\citet{fromang&pap07} who found without explicit dissipation a
monotonic decrease of $\alpha_{Max}$ as the resolution was 
increased. Here, no such systematic evolution is found as $Re$ goes
up. Indeed $\alpha_{Max}$ appears to vary only very weakly with
the Reynolds number. This is confirmed by the last column of
Table~\ref{model_prop} in which the values of $\alpha$, time--averaged
between $90$ and $130$ orbits, are
reported for the different models. The rate of angular momentum
transport appears to be somewhat smaller in model $Re6250$ than in
model $Re3125$ and $Re12500$. Nevertheless, the difference between the
three simulations remains less than $25 \%$. Taken together, the
three measurements suggest that $\alpha$ is of the order of $7 \times
10^{-3}$ and is fairly independent of the Reynolds number. At the
very least, a systematic evolution of $\alpha$ with $Re$ is ruled out
by the simulations.

%-Average value of alpha calculated between $t=90$ and $t=130$
%-Speak about $l_z$

%\section{Statistical properties}
%\label{stats_sec}

%In this section, three arguments are presented that suggest the
%existence of a well--defined injection length for the turbulence,
%independent of the Reynolds number.

\subsection{Power spectrum}
\label{power_spectra_sec}

\begin{figure}
\begin{center}
\includegraphics[scale=0.4]{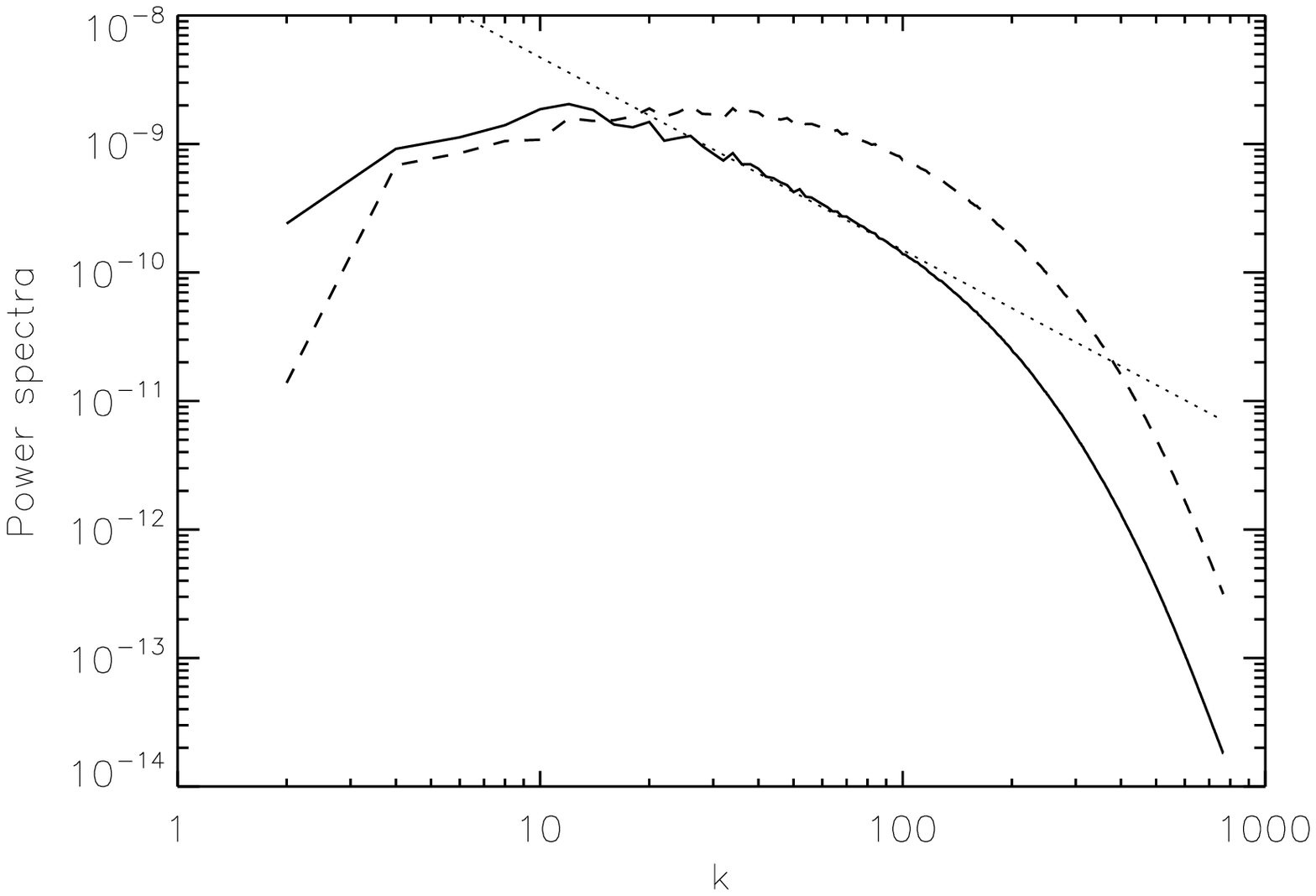}
\includegraphics[scale=0.4]{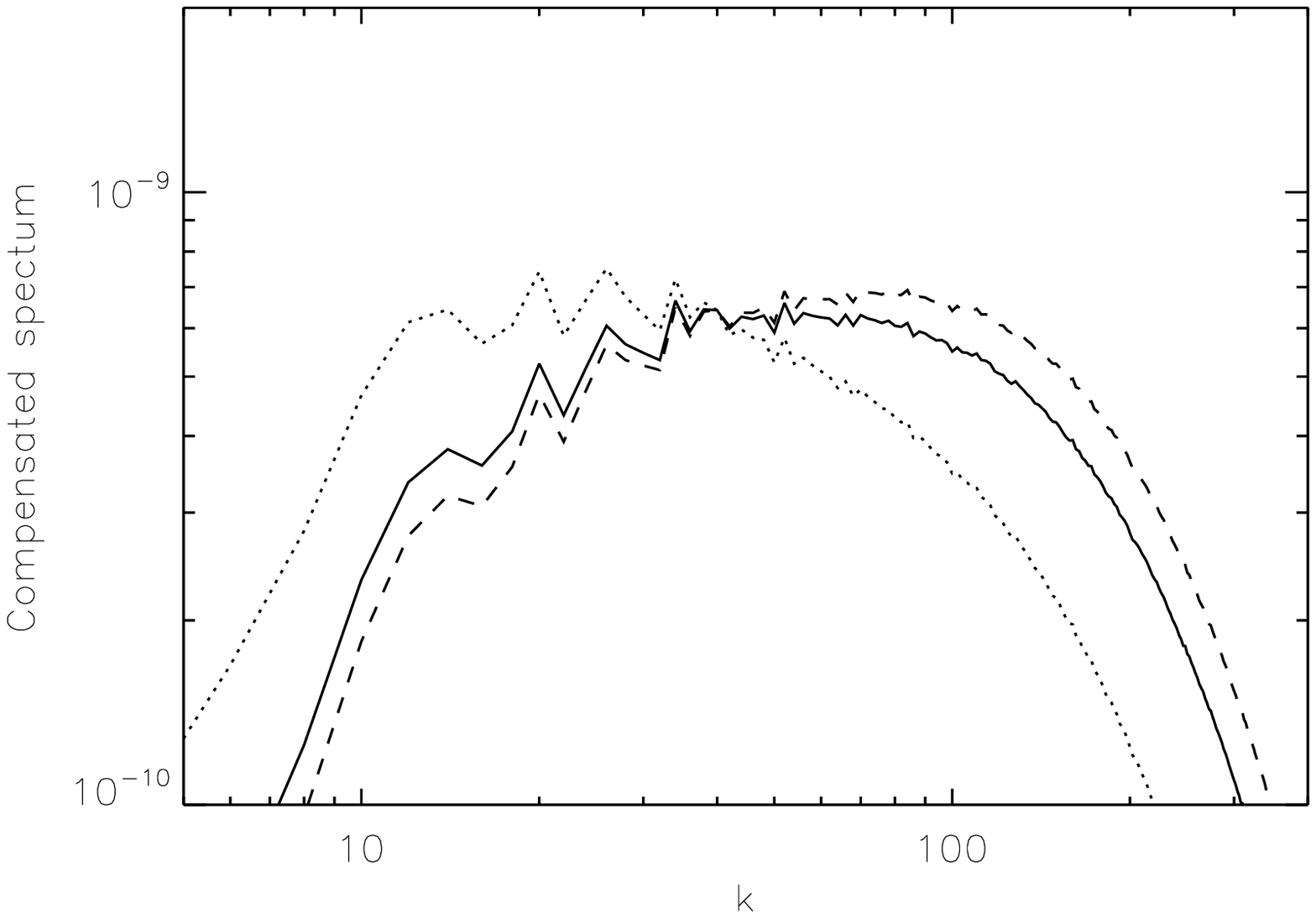}
\caption{{\it Top panel:} kinetic ({\it solid line}) and magnetic
  ({\it dashed line}) energy power spectrum for model $Re12500$, time
  averaged over twenty snapshots between $t=90$ and $t=120$. The
  dotted line shows a power law line with index $-3/2$ for the purpose
  of comparison. {\it Bottom panel:} kinetic energy power spectra
  compensated by $1$ ({\it dotted line}), $3/2$ ({\it solid line}) and
  $5/3$ ({\it dashed line}). Both panels are suggestive of a
  $k^{-3/2}$ spectrum in the range $30<k<100$.}
\label{power_spectra_Re12500}
\end{center}
\end{figure}

\begin{figure}
\begin{center}
\includegraphics[scale=0.4]{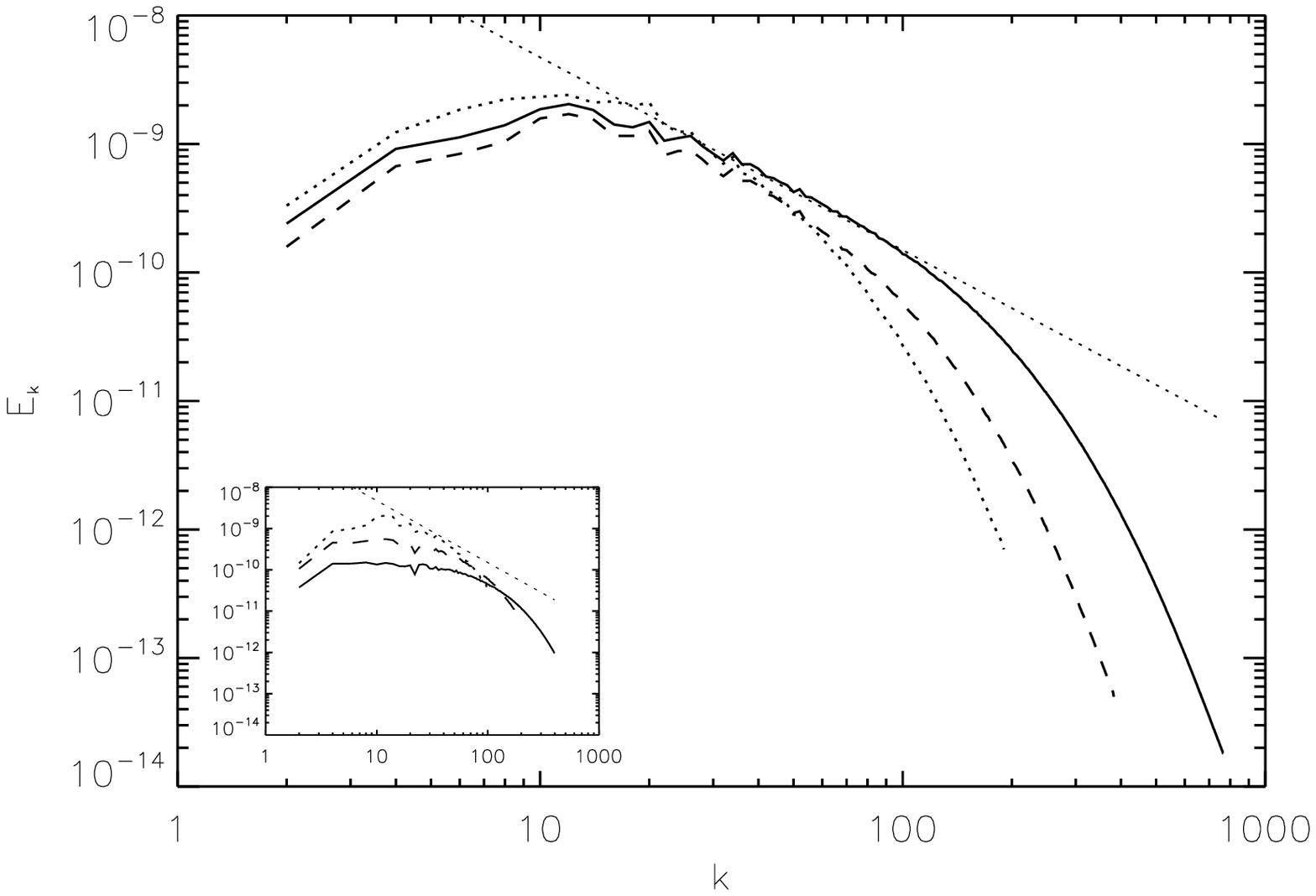}
\includegraphics[scale=0.4]{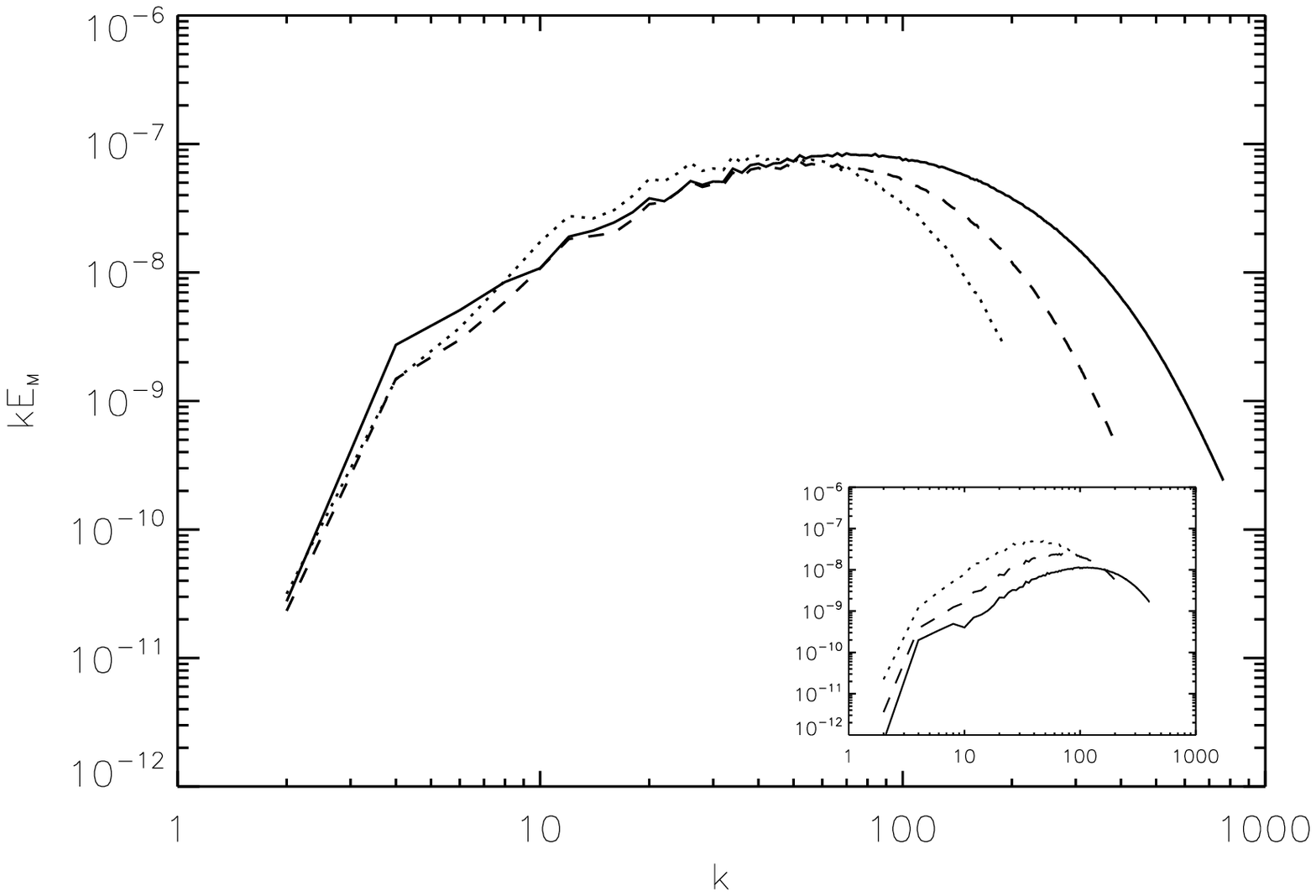}
\caption{{\it Top panel:} plot of $E_K$ for model
  $Re3125$ ({\it dotted line}), $Re6250$ 
  ({\it dashed line}) and $Re12500$ ({\it solid line}). The dotted
  line shows a power--law line with the index $-3/2$ for 
  comparison. As $Re$ and $Rm$ increase, the kinetic energy display an
increasing region well--fitted by a power law with the index $-3/2$, while
the viscous cut--off region moves to higher $k$ values. {\it Bottom
  panel:} Same as the top panel, but for the quantity $k E_M$. On both
panels the insets reproduce the results of
\citet{fromang&pap07} for model STD64 ({\it dotted line}), STD128
({\it dashed line}) and STD256 ({\it solid line}).}
\label{sequence_spec}
\end{center}
\end{figure}

The top panel of Fig.~\ref{power_spectra_Re12500} shows the
shell--averaged kinetic
energy power spectrum $E_K$ ({\it solid line}) and magnetic energy power
spectrum $E_M$ ({\it dashed line}) for model $Re12500$. The latter is
rather flat over about a decade in wavenumber (from $k \sim 10$ to $k
\sim 100$) and is larger than the kinetic energy over that range. By
contrast, the former displays a clear power--law behavior for
wavenumber $20<k<100$. For the purpose of comparison, the dotted
line shows a pure power--law with the index $-3/2$ that nicely fits the
solid line of the plot. By analogy with hydrodynamic turbulence it is
tempting to associate the large scale end of the power--law part of the
spectrum with an injection length $l_{inj} \sim 2\pi / k_{min} \sim
0.3 H$. Similarly, the small scale end can be associated with the viscous
cut--off length and is found to be $l_{visc} \sim 2\pi / k_{visc}
\sim 0.06 H$. This is  about $32$ cells at that
resolution and is thus well-resolved by the code. Furthermore,
results obtained in the kinematic regime of incompressible and homogeneous MHD
turbulence suggest that the resistive length $l_{res} \sim
Pm^{-1/2}l_{visc}$ \citep{schekochihinetal04}. Thus, $l_{res}$ is
of order $16$ cells and also well resolved, which shows that 
numerical dissipation is most likely negligible in this simulation. Given the
still limited resolution of model $Re12500$, the reliability of the
power--law exponent mentioned above can however be questioned: for
that purpose, the bottom panel of 
Fig.~\ref{power_spectra_Re12500} displays three compensated spectra
of $E_K$, $kE_K$ ({\it dotted line}), $k^{3/2}E_K$ ({\it solid
line}) and $k^{5/3}E_K$ ({\it dashed line}) respectively. First, the figure
illustrates the difficulty of a reliable determination of the
exponent. Indeed, the power--law extends over less than a decade in
wavenumber. Nevertheless, the dashed line, which unambiguously rises
over the interval of $10<k<100$, excludes a $k^{-5/3}$ spectrum and
rather suggests an exponent larger than $-5/3$. The dotted line on
the other hand suggests $-1$ as an upper limit. Finally, the solid
line suggests $k^{-3/2}$ as a tentative fit for the power--law
range of the spectrum ($30<k<100$). Finally, Fig.~\ref{sequence_spec} ({\it 
top panel}) compares the shape of $E_K$ in model $Re3125$ ({\it
dotted line}), $Re6250$  ({\it dashed line}) and $Re12500$ ({\it solid
line}). For all models, the kinetic energy power spectrum peaks at $k
\sim 10$--$20$. For larger wavenumbers, the $k^{-3/2}$ power--law
becomes more and more apparent as the Reynolds number 
increases. The bottom panel of Fig.~\ref{sequence_spec} plots the
quantity $kE_M$ for the three simulations. The peak of each curve thus
provides an estimate of the scale at which magnetic energy is
located. It is found to lie at $k_{peak} \sim 30$--$40$,
$50$--$60$ and $70$--$80$ respectively when $Re=3125$, $6250$ and
$12500$. In other words, the scale at which most of the magnetic energy is
located moves toward smaller and smaller scales as $Rm$ is
increased. This is different 
from the results reported by \citet{haugenetal03}, but not unexpected
given existing theories of small scale dynamos with large $Pm$
\citep{schekochihinetal02a,schekochihinetal02b}. On both panels,
the small insets plot the spectra obtained by \citet{fromang&pap07}
without explicit dissipation. Aside from the decrease of
their amplitude with resolution, the most noticable differences with
the results presented here are twofold: first, the kinetic
energy power--spectra appear flatter at intermediate
wavenumbers. In addition, there is more energy (both kinetic and magnetic)
at the smallest scales of the box.
%than for the runs with
%explicit dissipation that have been considered in the present letter.}

%For the Bfield case,
%careful to average over regions over which the field is about the
%same: 10 to 30 in model512, 20 to 39 in model256 but only 43 to 47 in model128...

\subsection{Correlation length}
\label{correlation_sec}

\begin{figure}
\begin{center}
\includegraphics[scale=0.22]{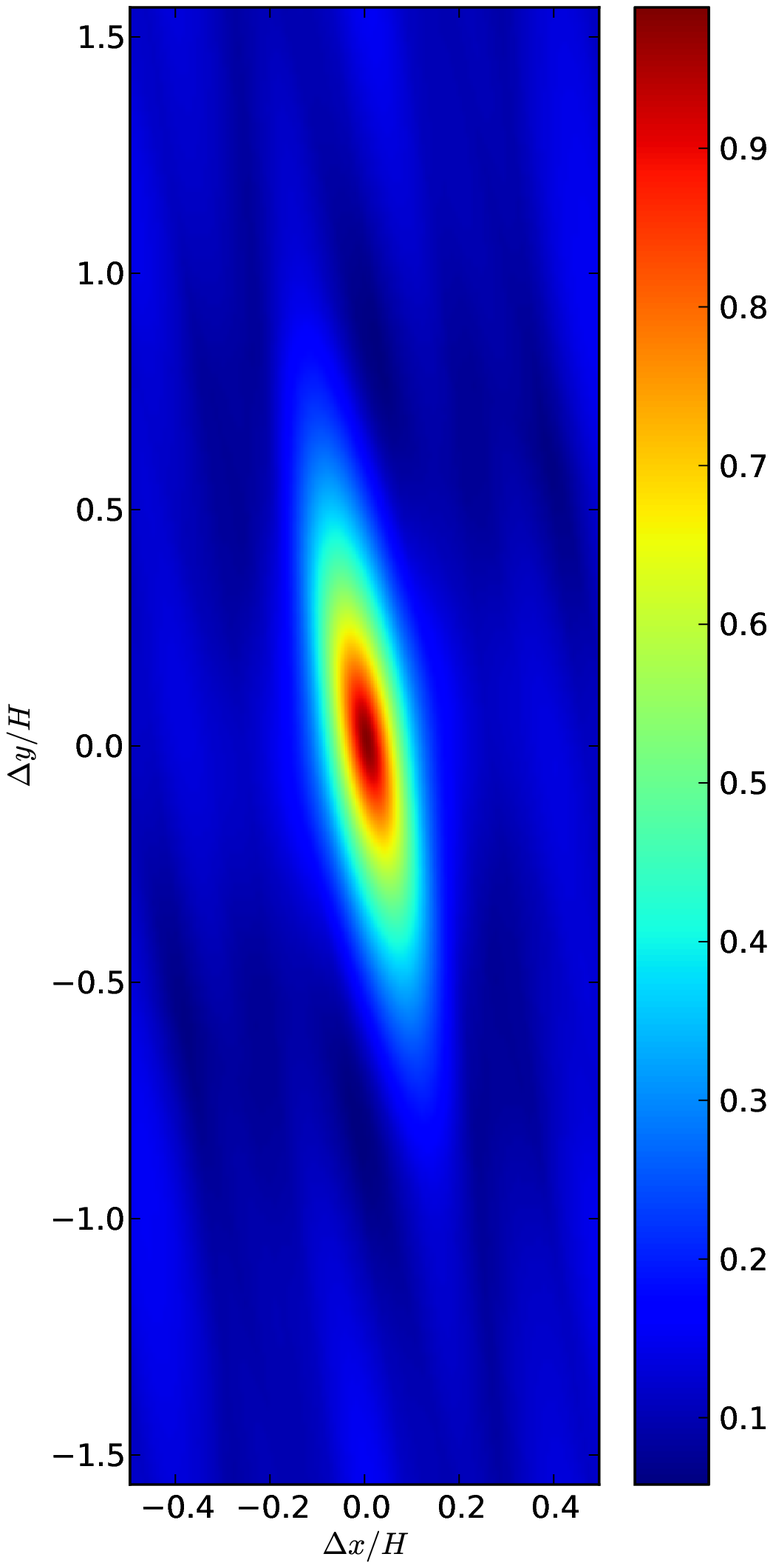}
\includegraphics[scale=0.22]{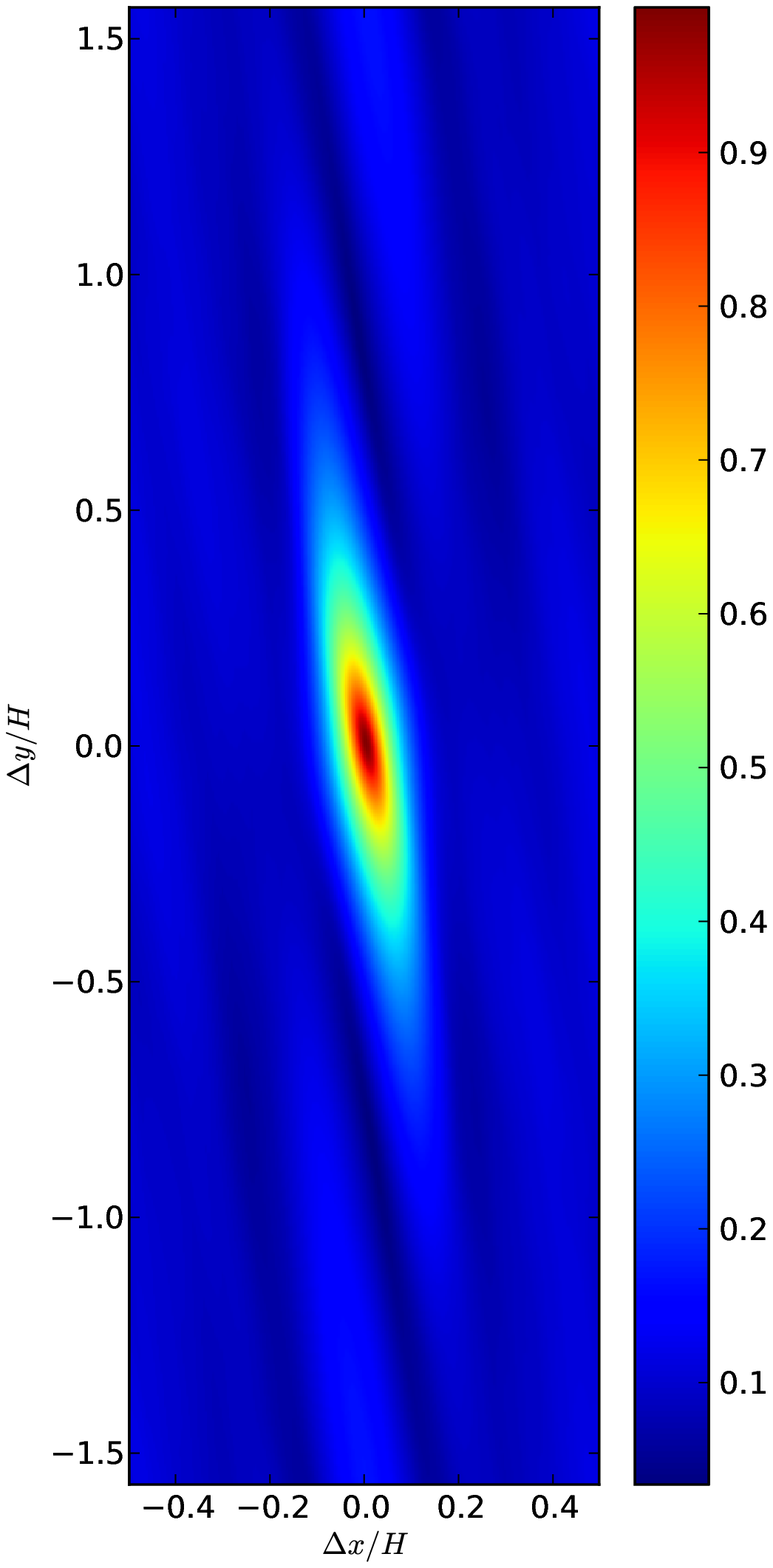}
\includegraphics[scale=0.22]{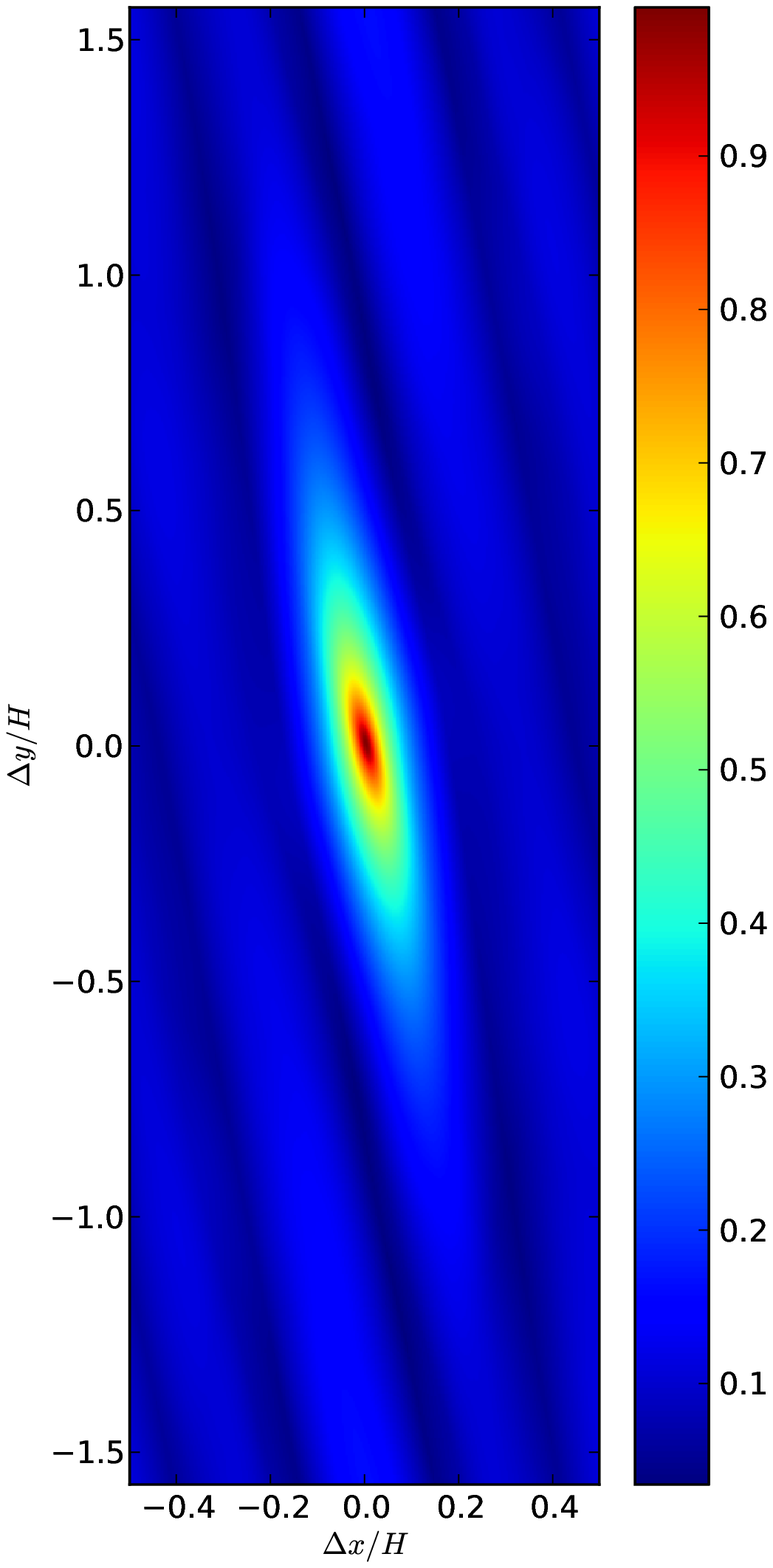}
\caption{Structure of the correlation function $\xi_v$ in the
  $(\Delta_x,\Delta_y)$ plane for model $Re3125$ ({\it left panel}),
  $Re6250$ ({\it middle panel}) and $Re12500$ ({\it right panel}). Its
  structure is only weakly dependent on the Reynolds number,
  suggesting a well--defined injection length.}
\label{sequence_correl}
\end{center}
\end{figure}

The shape of the kinetic energy power--spectrum described above
suggests an injection length $l_{inj}$ that appears to be independent
of $Re$ for the range of the Reynolds numbers investigated
here. However, the shell average involved 
in its derivation washes out all information about the anisotropy of
the turbulence.
% due to both shear and rotation
This can be
investigated using the two--points--correlation function
\citep{guanetal09,davisetal10}:
%. It is calculated according to 
\begin{equation}
\xi_v(\bb{\Delta x})= \textrm{$<$} \Sigma_i \delta v_i(\bb{x}) \delta
v_i(\bb{x}+\bb{\Delta x})\textrm{$>$}/ \textrm{$<$}\Sigma_i \delta
v_i^2 \textrm{$>$} \, .
\end{equation}
Here $<$.$>$ stands for a volume average, $\delta v_i(\bb{x})$
corresponds to the velocity fluctuations in the direction i and the
sum is over spatial coordinates.
%and there
%is an implied summation over the three spatial components of the
%velocity. 
%A similar definition is adopted for the magnetic field
%correlation function $\xi_B(\bb{\Delta x})$. 
Isocontours of $\xi_v$ in
the plane $\Delta z=0$ are shown in Fig.~\ref{sequence_correl}. From
left to right, the different panels correspond to models
$Re3125$, $Re6250$ and $Re12500$ respectively. As found by \citet{guanetal09} and
\citet{davisetal10}, $\xi_v$ has an ellipsoidal shape for all
models. The tilt angle $\theta_v$ of the major axis is $\theta_v$$\sim$$8$
degrees, which agrees with the results of \citet{guanetal09}. Following the
procedure outlined by these authors (i.e. by fitting the shape of the
correlation function in a given direction by an exponential), the
correlation lengths along the major, minor and vertical axis of the
ellipsoid are found to be 
$(\lambda_{min},\lambda_{max},\lambda_{z})=(0.08,0.45,0.08)H$, 
independently of the Reynolds number. This suggests once
more that the injection length of the turbulence is only weakly
dependent on the Reynolds number. At the same time, the small
difference between $\lambda_{min}$ or $\lambda_{z}$ and $l_{visc}$
as quoted in section \ref{power_spectra_sec} is a warning that the
injection range and the dissipative range might overlap in these simulations.

%V1
%. These
%three values are comparable in amplitude with the estimate obtained in
%section~\ref{power_spectra_sec} and are almost independent of the Reynolds
%number. Again, 

%Moreover, the values 
%quoted above for $\lambda_{min}$, $\lambda_{max}$ and $\lambda_{z}$
%bracket the estimate obtained in section~\ref{power_spectra_sec} using
%a {\it shell--averaged} kinetic energy power spectrum. The two
%estimates are thus consistent with each other.

%Velocity correlation function: For all models, we found $i \sim 8$
%degrees and the correlation lengths are:
%$(\lambda_{min},\lambda_{max},\lambda_{z})=(0.08,0.45,0.08)H$. Even
%for model128, $\lambda_{max} \sim \lambda_{z}$ corresponds to $10$ cells and for
%model512 it is about $40$ cells.

%Magnetic field: for model128, appear to have
%$(\lambda_{min},\lambda_{max},\lambda_{z})=(0.05,0.35,0.05)H$ when
%averaging over all snapshots and $i \sim 12$.

%\subsection{Total energy injection}
%
%\begin{figure}
%\begin{center}
%\includegraphics[scale=0.45]{figure/spec_S.ps}
%\includegraphics[scale=0.45]{figure/spec_D.ps}
%\caption{}
%\label{power_spectra_pm4}
%\end{center}
%\end{figure}

\section{Conclusion}
\label{conclusion_section}

Here zero net flux high resolution numerical simulations of MRI--driven MHD
turbulence are used to demonstrate this result: when
$Pm$$=$$4$, the dependence of $\alpha$ on the Reynolds number is very
weak. In all models, $\alpha \sim 7 \times 10^{-3}$ to within about
$15\%$. This result unambiguously shows that the decrease of 
$\alpha$ with 
resolution reported by \citet{fromang&pap07} is a numerical artifact
that contains no physical information about the nature of the MHD
turbulence in accretion disks. Quite differently, the present
simulations are consistent with a nonzero value of $\alpha$ at
infinite Reynolds numbers for a magnetic Prandtl number
higher than unity. Note that this weak dependence of $\alpha$ with
$Re$ for $Pm>1$ is also suggested by the data recently reported by 
\citet{simon&hawley09} and \citet{longaretti&lesur10} respectively for
a mean azimuthal and vertical magnetic field.

In addition, a number of statistical properties of the turbulence are
reported. The kinetic energy power spectrum of the turbulence and the
two--points--correlation function of the velocity both suggest a
well--defined injection length $l_{inj}$ of a few tens 
of a scaleheight. For the range of the Reynolds numbers $Re$ that can be
probed with current resources, $l_{inj}$ seems to be
independent of $Re$. At the highest resolution achieved here, the
kinetic energy power spectrum displays a power--law 
scaling over almost a decade in wavenumber. However, given the limited
extent of the power--law range, the precise exponent of this
power--law cannot be accurately determined: an exponent of $-3/2$ appears
to be consistent with the data, while a $-5/3$ exponent seems too
steep. Nevertheless, as suggested in Sect.~\ref{correlation_sec}, the 
separation between the forcing and the dissipative scales might still
be marginal. This is why a detailed comparison of these exponents with 
existing MHD turbulence theories
\citep{iroshnikov63,kraichnan65,goldreich&sridhar95} 
is probably premature at this stage. Higher resolution
simulations are definitively needed. Finally, the shape of the
magnetic energy power spectrum shows that magnetic energy is
mostly located at small scales and shifts to smaller and smaller
scales as $Rm$ increases, as expected from small scale dynamo theory
\citep{schekochihinetal02a}. This is consistent with the scenario
postulated by \citet{rinconetal08} of a large scale MRI forcing
that generates and coexists with a small scale dynamo.

%Finally, the limitations of this work should be mentionned. They are
%due both to the limited realism of the present simulations (no density
%stratification, limited range of $Pm$ and $Re$ that can be probed) but
%are also of computational nature: in the future, the resolution will
%have to be increased further and the simulations to be run longer for 
%stonger conclusions about the nature of MRI--induced MHD turbulence to
%be drawn.

%Of course, there are many limitations to these results: first and
%foremost, the resolution, albeit at the limit of current computational
%resources, is still too small for definite statements to be
%made. Second, the simulations are largely idealized. In particular,
%the neglect of density stratification may change the picture described
%in this letter by introducing new effects at larger scales than those
%described here. 

%Of course, there are many limitations to these results. Density
%stratification, resolution, are all important aspects of this work
%that will have to be improved and/or included in the future for
%definite conclusions about the nature of MRI--induced MHD turbulence
%to be drawn. 

\section*{ACKNOWLEDGMENTS}

The author acknowledges insightful discussions with F.~Rincon, G.~Lesur
and P.--Y.~Longaretti and is indebted to S.~Pires for her help in
analyzing the data presented here. These simulations
were granted access to the HPC resources of CCRT under the allocation
x2008042231 made by GENCI (Grand Equipement National de Calcul Intensif).

\bibliographystyle{aa}
\bibliography{author}

\end{document}